\documentstyle[psfig,aps,twocolumn]{revtex}

\begin{document}
\draft
\title{%{\rm Submitted to American Journal of Physics}\hfill\null\\[5mm]
Discreteness effects on soliton dynamics: a simple experiment.}
\author{Claude Laroche, Thierry Dauxois and Michel Peyrard}
\address{Laboratoire de Physique, UMR-CNRS 5672,\\
Ecole Normale Sup\'erieure de  Lyon, \\
 46 All\'{e}e d'Italie, 69364 Lyon C\'{e}dex 07, France}
\date{\today}
\maketitle
\begin{abstract}
We present a simple laboratory experiment to illustrate some aspects
of the soliton theory in discrete lattices with a system that models
the dynamics of dislocations in a crystal or the properties of
adsorbed atomic layers.  The apparatus not only shows the role of the
Peierls-Nabarro potential but also illustrates the hierarchy of
depinning transitions and the importance of the collective motion in
mass transport.
\end{abstract}

\bigskip

Since the discovery of solitary waves by J. Scott Russell~\cite{RMM}
on his horse along the Edinburgh canal, and the seminal paper by
Zabusky and Kruskal~\cite{zabuskykruskal} explaining the recurrence
phenomenon found by Fermi-Pasta-Ulam in a nonlinear chain of atoms,
the soliton concept has been the starting point of very fruitful
researches: it has not only lead to very important advances in applied
mathematics but it is also an excellent tool to explain the properties
of various physical systems~\cite{LA}.

The concept of soliton is generally associated with continuous fields,
and for instance the very interesting resource letter on solitons,
recently published in the American Journal of Physics
\cite{ResLettSol}, does not contain more than a few lines on discrete
lattices in spite of the fact that many physical systems are
intrinsically discrete, such as crystals or macro-molecules. 
This is because, except for very few specific cases such as Toda
lattice, the discretized version of integrable partial differential
equations which describes physical systems
does not have true soliton solutions
 However this does not mean that the
soliton concept is useless to describe nonlinear excitations in
discrete systems. Such systems can sustain approximate soliton
solutions, and discreteness introduces new features such as soliton
pinning or radiation of small amplitude waves. These effects are
important to understand some physical phenomena. For instance it is
because their width is of the same order as the lattice spacing that
dislocations do not propagate freely in a crystal and radiate
vibrational modes when they are forced to propagate by an external
stress. This explains the familiar observation that, when one
successively bends and unbends a piece of iron wire many times it gets
hot. Each plastic distortion moves dislocations that radiate
vibrational waves in the solid, rising its temperature.

The present paper has a dual purpose. First we want to give a brief
tutorial on quasi-solitons in discrete systems using the example of
the diffusion of atoms on a crystalline surface because, on one hand
it allows a rather intuitive understanding of the phenomena and, on the
other hand it shows how collective effects can lead to new
phenomena. Then we introduce a simple experiment, easily build in a
physics laboratory, which illustrates
some of the discreteness effects that cannot be shown with the
soliton experiments which have been proposed earlier, such as water
waves experiments (see review by Segur in Ref.~\cite{LA}) or the 
chain of strongly coupled pendula proposed by A.C. Scott
~\cite{scott}. 

\bigskip
\section{ A brief tutorial on collective diffusion.}

Before describing the apparatus and experimental results, 
let us recall some basic ideas on the
diffusion of particles over a periodic potential under the effect of
an applied force.  The motion of a single particle can be easily
analyzed if one considers only the low temperature situation where the
thermal fluctuations can be neglected. In the absence of a driving
force, the particle is trapped into a potential well. Applying a
driving force is equivalent to tilting the potential and, above a
critical angle $\alpha_0$, the minimum disappears and the particle starts
to  slide on the washboard potential~\cite{Risken1984}.

Instead of a single particle let us now consider  a chain of particles
coupled by harmonic springs with an equilibrium length equal to the
distance between the potential minima: this state is called a
commensurate state. Its ground state is reached
when  all the particles are in the potential
minima (fig.~\ref{fsol}.a) and the situation is 
comparable to the case of a single
particle: the motion induced by an external force will only start when
the minima of the total potential (including the force term) have
disappeared. 
If a defect is introduced in the chain, the situation becomes very
interesting and is equivalent to creating a dislocation in a
crystal. Consider for instance the case of a missing particle obtained
by moving one half of the chain by one unit, i.e. a kink in the
particle positions.  In the vicinity of the defect, the competition
between the elastic energy of the springs and the substrate potential
energy displace the particles with respect to the minima of the
potential (fig.~\ref{fsol}.b).  
As a result the particles next to the defect are easier to
move with an external force than a particle sitting at the bottom of
the well.
For a sinusoidal potential of period 
$a$, $V(x) = V_0 [1 - \cos(2 \pi x / a)]$, when the coupling between
the particles is strong, the position of the n$^{th}$
particle is $x_n=a [n+{2\over \pi} \tan^{-1} \exp(n/\ell)]$, with
$\ell = {a\over 2\pi} \sqrt{k / V_0}$ where  $k$ is the spring elastic constant
\cite{scott}. No analytical solution for the structure of the defect
is known in the case of weak coupling.

When the coupling between the particles is very {\em strong}, the
 defect is extended. The displacements of the particles vary
 progressively from zero to one lattice spacing across the defect. 
Consequently, there are particles at any level of the
 substrate potential, including on the maximum. This is the situation
 that is realized in the chain of strongly coupled pendula described
 by Scott~\cite{scott}. In this quasi-continuum situation, it is easy
 to understand why the defect (or soliton in the terminology of
 nonlinear science) can move freely. When it is translated some
 particles have to climb over the substrate potential barriers but
 simultaneously others move downward in the potential and the overall
 translation does not require any energy.  In a continuum model, the
 system is invariant by {\em any} translation and the free motion of
 the soliton is simply a manifestation of this translational symmetry
 (Goldstone mode).

In the {\em weakly} coupled situation that we consider here, the
situation is different. The defect is highly localized and even though
some particles are slightly displaced from the potential minima
as in Fig.~\ref{fsol}(c), the
springs are not strong enough to maintain  particles at the top of
the substrate potential barriers.  Now, in order to translate the
defect, one has to move particles up on the substrate potential: there
is a barrier to the free translation of the defect, which is well known
in dislocation theory as the Peierls-Nabarro (PN) barrier. The weak
coupling case appears therefore as a natural intermediate case between
the case of individual particles which must overcome the full
potential barrier of the substrate potential and the continuum case
where the soliton is completely free to move. In this intermediate
case, the defect moves as a {\em collective excitation} over an {\em
effective potential}, the PN potential, which has the period of the
lattice and an amplitude which is much lower than the individual
potential barrier of the substrate.  This analysis in terms of a
collective object explains why the critical stress for the plastic
deformation of a crystal is several orders of magnitude lower than the
stress that would translate a full atomic plane above
another~\cite{kittel}. 

The case of atoms adsorbed over an atomic surface is even more
interesting because usually the equilibrium distance $b$ that the
atoms would select if they were free is not commensurate with the
period $a$ of the crystalline substrate. As a result the interaction
forces compete with the substrate potential to determine the particle
positions. The atomic layer minimizes the energy by letting the
particles drop near the bottom of the substrate potential wells almost
everywhere and compensating for the mismatch between $a$ and $b$ by
creating local discommensurations which are very similar to the
isolated defect described above for the commensurate case. The only
difference is that, depending on the commensurability ratio, there is
a hierarchy of defects with different shapes and different barriers.
In the case of an irrational ratio, this hierarchy is complete and
some defects have a vanishingly small PN barrier: in the discrete
lattice (provided the coupling is not too weak~\cite{aubry}), a
vanishingly small external force can cause mass transport in the
system.  The application of an external driving force to an
interacting chain of atoms exhibits this {\em hierarchy of depinning
transitions}~\cite{nlmobility}. For very low force, no motion can be
detected. Then, the geometrical kinks (the discommensurations due to
the concentration of particles related to the number of minima of the
potential) start to move whereas the atoms stay static.  For larger
external forces, additional defects ({\em kink-antikinks pairs}) are
created giving rise to an increase of the mobility.  Finally, for high
enough forces, all atoms are moving with the mobility of single
Brownian particles.

\section{Experiment}

Let us now describe the experimental apparatus sketched in
Figs.~(\ref{description}).  The system under study is a chain of steel
cylinders, each one 60 mm in length and 8 mm in diameter.  The
cylinders sit on a washboard potential cut in a block of plastic
(approximate sine-shape potential).  The height of the valleys is 10
mm and the lattice spacing is 20 mm. The cylinders are coupled one to
an other with an elastic string; the first cylinder is fixed whereas
the other ones are free to move. In the example shown in
Figs.~(\ref{description}), a 3 mm slot was made along the center of
the plastic support (represented by the two dotted lines in
Fig.~(\ref{description}b)) in order to let the elastic move freely.

The concentration of cylinders, which determines the presence and
structure of the discommensurations, is defined as
$\theta=M/N$ where $M$ is the number of cylinders and $N$ the numbers
of lattice spacing. The defects shown in Fig.~\ref{fsol}.b and
~\ref{fsol}.c  correspond
to $\theta = 15/16$ while figure \ref{description} shows schematically
the case $\theta=2/3$.
In the experiment the length $b$ of the elastic strings is adjusted
for each concentration by imposing the condition 
$\theta=a/b$ which means that, in the absence of the substrate
potential the cylinders would be equally spaced in such a way that $M$
cylinders would cover $N$ lattice spacings, achieving the desired
concentration.

The apparatus can be used to test the static and dynamical properties
of this model system. First one can measure the depinning force which
is required to move a lattice with a given concentration above the
substrate. This is done by progressively inclining the system,
i.e. increasing very slowly the angle $\alpha$ and
determining 
the critical angle $\alpha_c$ above which the initial distribution of
cylinders is unstable, i.e. above which at least one cylinder begins
to slide. The results of
this experiment are shown on Fig~\ref{angle}.  For $\theta=1/q$ (with
$q$=1,2,..), the system has a trivial ground state with one cylinder
at the bottom of the substrate potential wells every $q$ wells.
In these cases, all cylinders start to move simultaneously.
As discussed above these commensurate cases should be the hardest to 
depin and this is confirmed by the experiment. Moreover the behavior
should not depend on the number of empty wells that might separate two
cylinders, i.e. we expect the same critical angle for $q=1$ or $q=2$..
This is confirmed by the results shown on Fig~\ref{angle}. 
 When $\theta=p/q$ is a rational number with $ q > p$, $q \not= 1$,
such as $\theta=2/3$ shown in
Figs.~(\ref{description}), the ground state involves defects and
Fig~\ref{angle} shows that their cooperative motion (cases
$\theta=2/3$, $\theta=3/4$) occurs for lower angles than the
individual motion. This illustrates therefore the depinning hierarchy,
the Peierls Nabarro barrier being lower than the substrate barrier.
Figure \ref{angle} also shows that the PN barrier depends on the
commensurability ratio which governs the structure of the
kink-defects~\cite{bklast,nlmobility}; higher order rational numbers
result in a lower PN barrier:
$\alpha_c(2/3)>\alpha_c(3/4)\Rightarrow E_{\hbox{PN}}(2/3)>E_{\hbox{PN}}(3/4)$. With a system as small as the one we are using we cannot
study other rationals such as 3/5, 5/8, 8/13 that should lead to lower
barriers. A truly incommensurate case, leading to a vanishing PN
barrier~\cite{aubry}, corresponds to an irrational
ratio and therefore it cannot be obtained in an experiment since $M$
and $N$ are necessarily integers. This case can however be
approached by rational numbers with numerators and denominators chosen
in a Fibbonacci sequence~\cite{PAubry}, but it would require a model
much longer than the one we have built.

The second class of experiments that can be performed tests the
dynamical properties of the system by measuring the mobility of the
defect as a function of the applied force. This is done 
by artificially holding
the defect above the critical angle $\alpha_c$ while the potential is 
tilted, 
and then letting it go. When the
constraint is released, the defect slides on the washboard
potential. Using a high speed CCD camera to record the fast motion, we
measure the time $\Delta t$ for the propagation of the defect over $n$
lattice-spacings. 

Since the velocity is $\langle v\rangle={n a/\Delta t}$ 
and the external force is due
to gravity, $F=mg \sin(\alpha)$, the mobility is by definition
\begin{equation} 
B(\alpha)=\langle v\rangle/F={n a/\Delta t \over mg \sin(\alpha)}\quad.
\end{equation} 
The results, obtained for the concentration $\theta=2/3$ sketched in
Fig.~\ref{description},  are plotted in Fig.~\ref{mobility}: they
clearly show a plateau corresponding to the kink-running state,
obtained when only the defect moves,
leading to the first contribution to mass transport.
The final transition to the sliding state, where all cylinders
slide on the washboard potential, is reached for higher external
forces.

\section{Conclusion.}

In this brief report, we have presented a simple teaching experiment
stressing the kink-concept in a discrete system.  One is able to
illustrate different theoretical ideas in a simple way using this
experiment.  The Peierls-Nabarro potential, usually presented in the
context of dislocation theory, and its role are not only clearly
emphasized but one shows that it is a function of the
concentration~\cite{aubry}, i.e. for atoms adsorbed on a crystal, it
varies with the coverage.  Moreover, this experiment explains the
recently developed idea about two-dimensional diffusion of
atoms~\cite{nlmobility}.  One easily detects a hierarchy of
depinnings: first the defects (kinks or antikinks) are moving, then
for higher forces, we have the individual motion.  The present
apparatus is too short to illustrate experimentally the existence of
an hysteresis phenomenon in the force driving the
diffusion~\cite{nlmobility}: the chain starts to slide for a force
$F_1$ but stops only when the force has been lowered below $F_2 < F_1$
because, once the motion has been initiated, the kinetic energy allows
the particles, or the defect, to overcome a small potential barrier.
It would be very interesting to build a much longer chain, so that few
defects could coexist and of course interact.  The critical angle of
the different concentration regions would also be different.

This experiment has however some important differences with the
physical problem of atomic diffusion. The elastic strings apply a
force to the cylinders only when they are extended (not in
compression) and the unavoidable solid friction has no simple
equivalent at the microscopic level. It is therefore important not to
overemphasize these results. Nevertheless, as recent very nice
experiments in Josephson junction arrays~\cite{ustinov} have confirmed
discreteness effects on soliton-like structures (see review by Peyrard
in Ref.~\cite{LA}), we think that the present experiment, conceptually
and materially more appropriate for teaching purpose, could be an
excellent tool to present the soliton concept in the framework of
discreteness to non-specialists.

\begin{figure}
\caption{Positions of a set of harmonically coupled
  particles subjected to a periodic potential. The springs connecting
  the particles (schematized by the thick line) have an equilibrium
  length equal to the period of the potential. \\
(a) Ground state of the system. All particles are in the potential
  minima. $\theta=1$. \\
(b) Excited state. The particles on the right half of the chain have
  been moved to the next potential well, creating a localized defect
  in the particle distribution. $\theta=1-1/N$.\\
(c) The same excited state for a weaker coupling $k$ (weaker connecting
  springs). The defect is narrower (i.e. the width $\ell$ is smaller) 
and the atoms in the defect core
  are closer to the minima of the potential. The depinning force
  required to  move the defect (c) is therefore higher that the
  depinning force in case (b).$\theta=1-1/N$.}
\label{fsol}
\end{figure}

\begin{figure}
\caption{Sketch of the experiment in the case $\theta=2/3$.  The side
view is presented in (a), whereas (b) presents the top view. In (a),
the arrow $g$ indicates the direction of the gravity field. The top
arrow means that this cylinder is always kept fixed. In Fig. (b), the
dotted lines represent the slot made in the plastic support in order
to let the elastic, represented by the solid line, free.}
\label{description}
\end{figure}
\begin{figure}
\caption{Critical angle above which the initial concentration is unstable,
giving rise to a moving kink-soliton. The concentration is defined as
the ratio between the number of cylinders and the number of sites.}
\label{angle}
\end{figure}

\begin{figure}
\caption{Mobility in the case $\theta=2/3$, represented in 
Fig.~{\protect{\ref{description}}},
versus the external force, monitored
by changing the angle $\alpha$. The results illustrate the three
different regions, and the hierarchy of depinning transitions.}
\label{mobility}
\end{figure}

\vfill\eject\null\vfill\eject
\psfig{figure=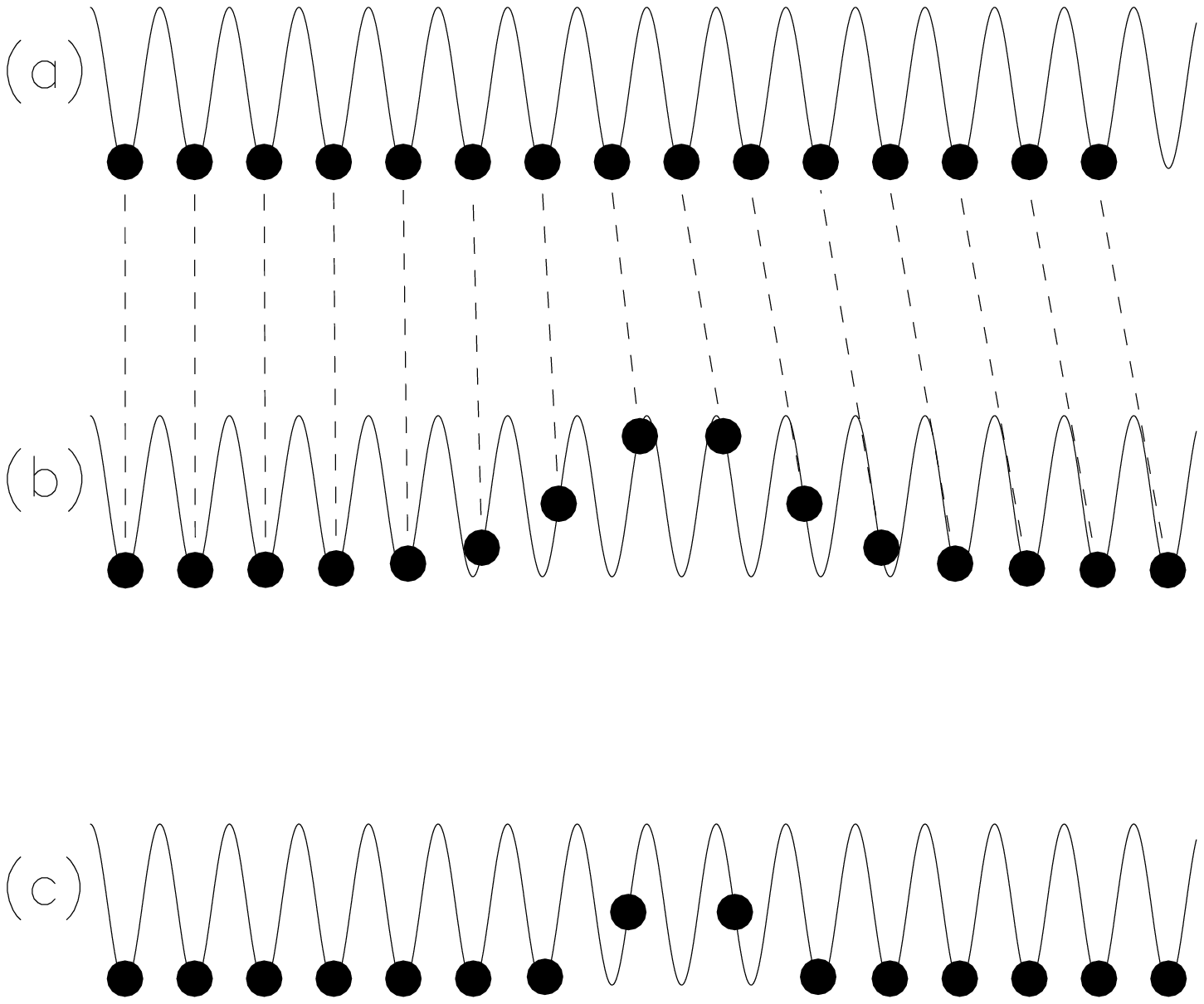,width=17cm}
\centerline{ Fig.~\ref{fsol}}
\vfill\eject\null\vfill\eject
\psfig{figure=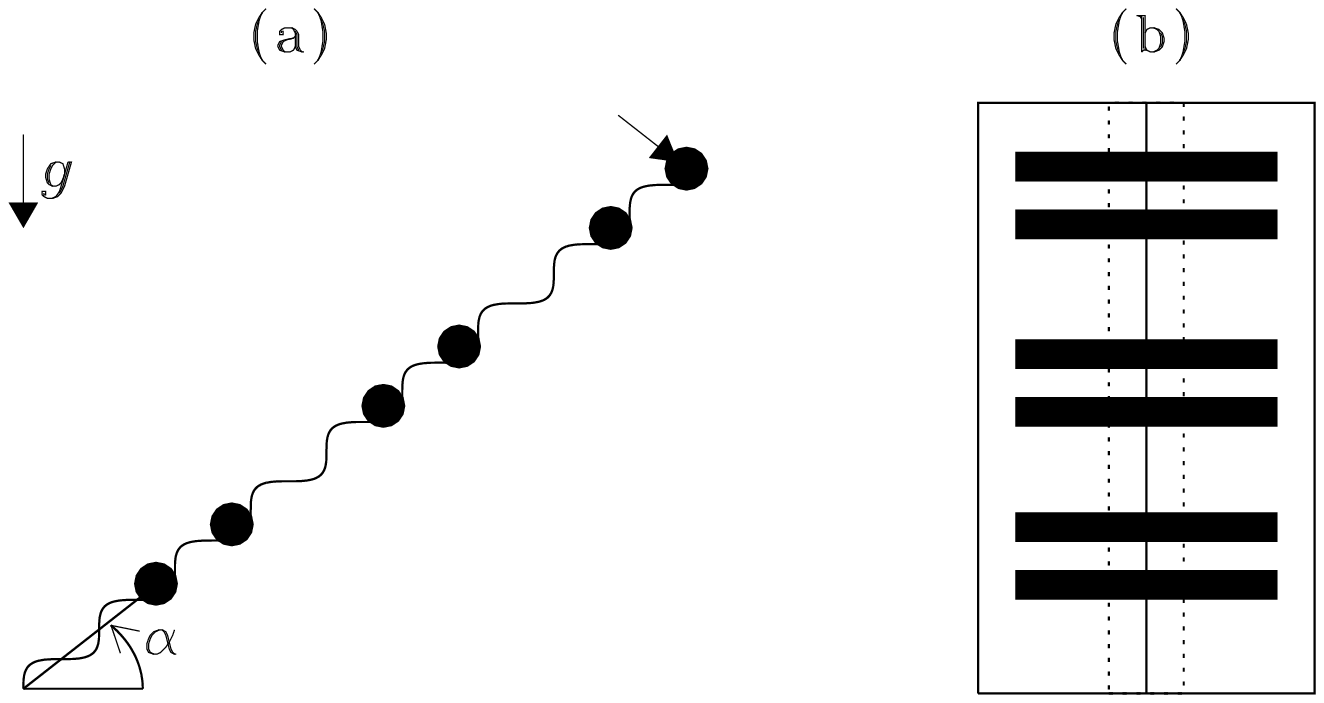,height=10truecm,width=19truecm}
\centerline{ Fig.~\ref{description}}
\vfill\eject\null\vfill\eject
\psfig{figure=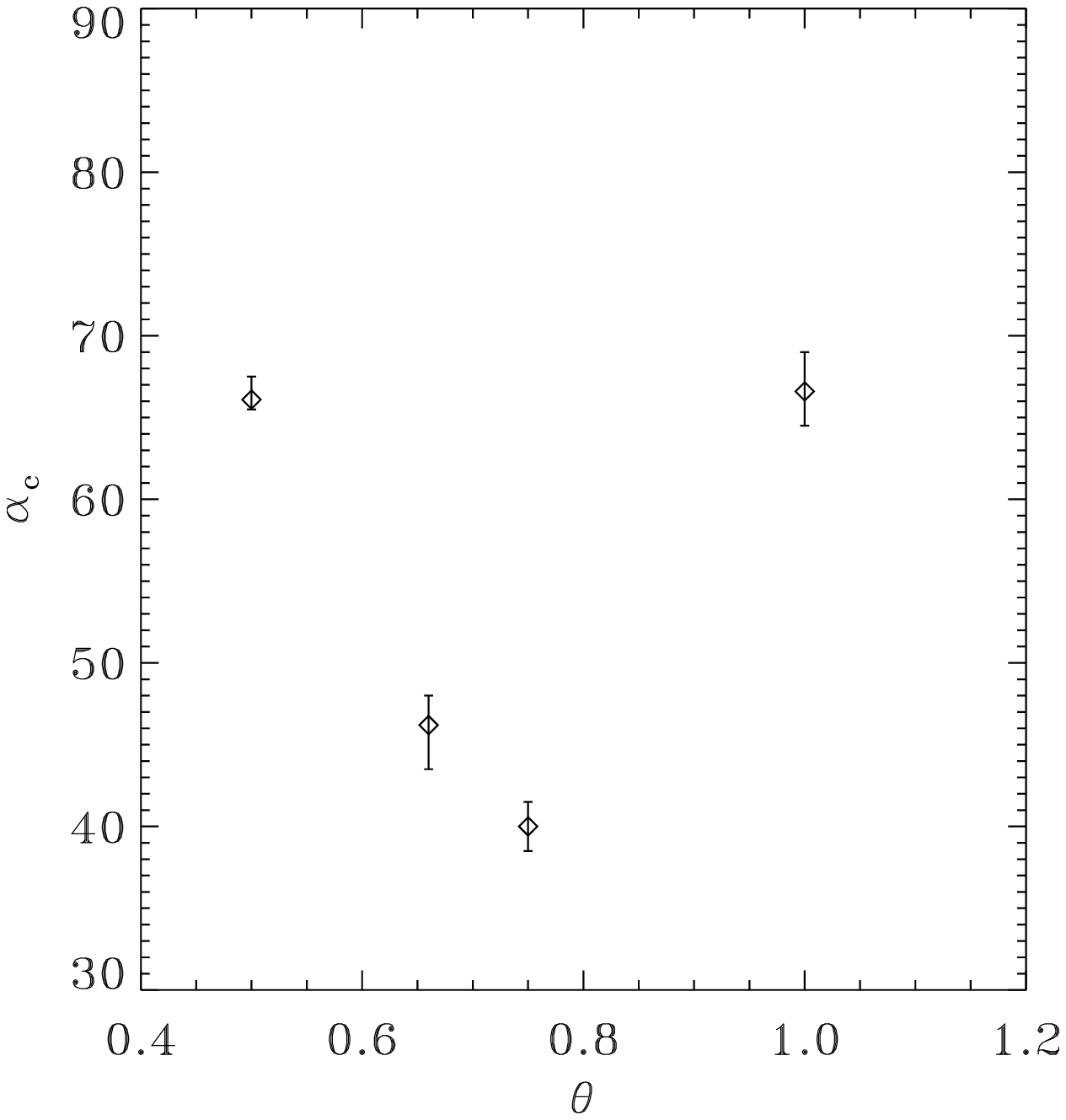,height=15truecm,width=15truecm}
\centerline{ Fig.~\ref{angle}}
\vfill\eject\null\vfill\eject
\psfig{figure=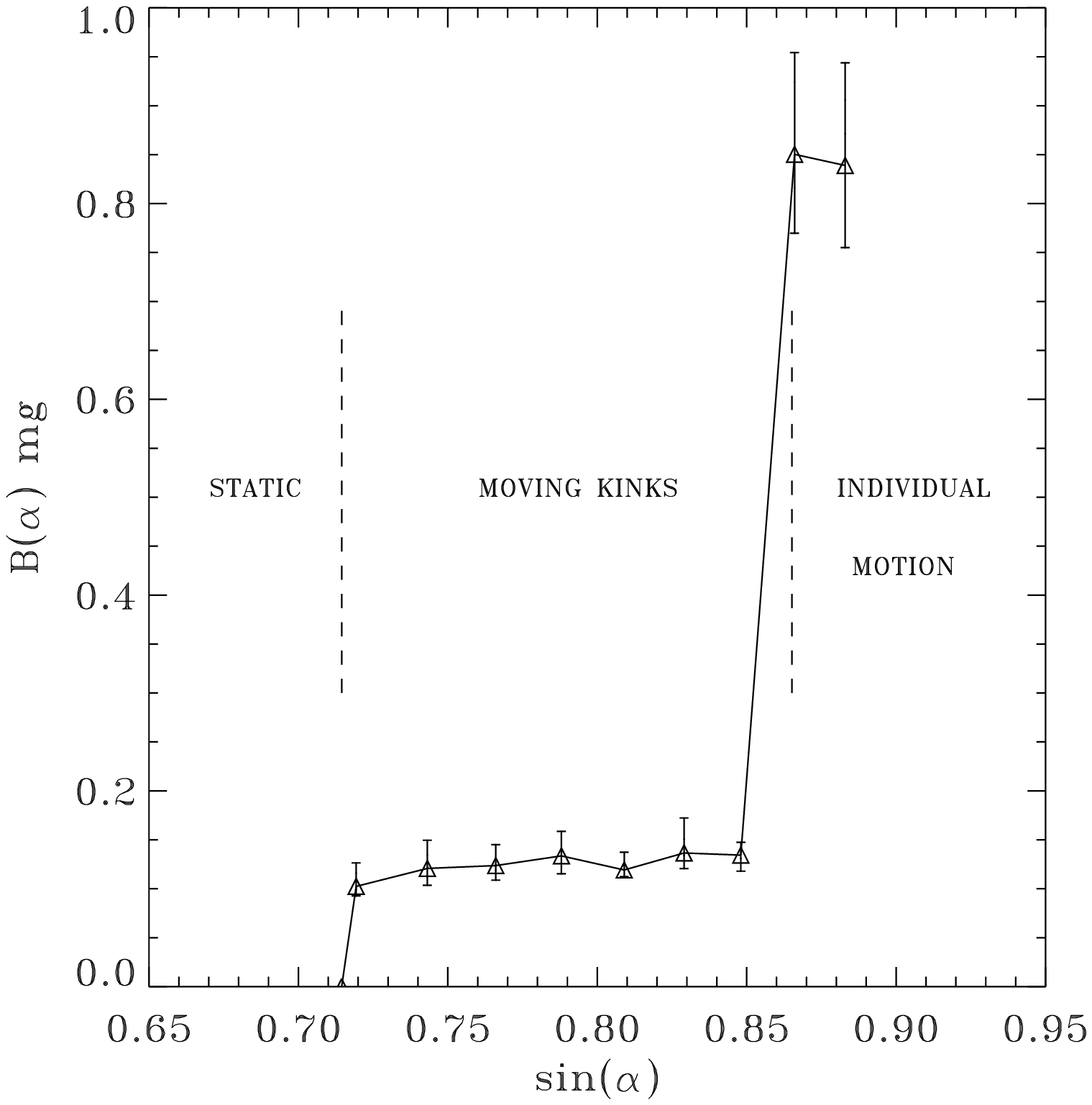,height=15truecm,width=15truecm}
\centerline{ Fig.~\ref{mobility}}

\end{document}